\documentclass[conference]{IEEEtran}

\usepackage{algorithm,algorithmicx}
\usepackage{algpseudocode,empheq,hhline}
\usepackage{dsfont}
\usepackage{amsmath,amsthm,amssymb,url,graphicx,rotating,ifthen,epsfig,array,color}
\usepackage{float}
\usepackage{dsfont}
\usepackage{mathtools}
\usepackage{multirow}
\usepackage{mathtools}
\usepackage{enumitem}
\usepackage{soul}
\usepackage{calligra}
\usepackage{cite}
\usepackage{rotating,longtable}
\usepackage[round]{natbib}
\usepackage{balance}
\usepackage{flushend} 
\usepackage{subcaption}
\usepackage{authblk}

\def\R{{\mathbb R}}
\def\E{{\mathbb E}}
\def\N{{\mathbb N}}
\def\e{\epsilon }
\def\K{{\cal K}}
\def\S{{\cal S}}

\def\T{{\cal T}}
\def\w{\omega}
\newcommand{\vect}[1]{\mathbf{#1}}
\begin{document}

%

%


\title{Scenario-based decision-making for power systems investment planning}

\author[1,2]{Jialin Liu}
\author[1,3]{Olivier Teytaud}
\affil[1]{Inria TAO, LRI, UMR 8623 (CNRS - Univ. Paris-Saclay)\\Gif-sur-Yvette, France}
\affil[2]{CSEE, University of Essex\\Colchester, United Kingdom}
\affil[3]{Google\\Z\"urich, Switzerland}
\affil[ ]{\textit{jialin.liu@essex.ac.uk}}
\maketitle
%

\begin{abstract}
The optimization of power systems involves complex uncertainties, such as technological progress, political context, geopolitical constraints. Negotiations at COP21 are complicated by the huge number of scenarios that various people want to consider; these scenarios correspond to many uncertainties.
These uncertainties are difficult to modelize as probabilities, due to the lack of data for future technologies and due to partially adversarial geopolitical decision makers.
Tools for such difficult decision making problems include Wald and Savage criteria, possibilistic reasoning and Nash equilibria.
We investigate the rationale behind the use of a two-player Nash equilibrium approach in such a difficult context; we show that the computational cost is indeed smaller than for simpler criteria.
Moreover, it naturally provides a selection of decisions and scenarios, and it has a natural interpretation in the sense that Nature does not make decisions taking into account our own decisions. 
The algorithm naturally provides a matrix of results, namely the matrix of outcomes in the most interesting decisions and for the most critical scenarios. These decisions and scenarios are also equipped with a ranking.
\end{abstract}

\section{Introduction: decision making in uncertain environments}\label{sec:rdm}

Planning in power systems relies on many uncertainties.
Some of them, originating in nature or in consumption, can be tackled through probabilities~\citep{RTEForecast,pinson2013renewable,siqueira2006stochastic,Vassena2003}; others, such as technology evolution, geopolitics or CO2 penalization laws, are somewhere between stochastic and adversarial:

{\bf Climate:} The United Nations Climate Change Conference, COP21, aims at achieving a new universal agreement on climate agreement, which is an issue of cooperation and competition.

{\bf Uranium supply:} India has been using imported enriched uranium from Russia since 2001. In 2004, Russia deferred to the Nuclear Suppliers' Group and declined to supply further uranium for India's reactors. The uranium supply was not resumed until the end of 2008 (after the refurbishment was finished). Now, Russia is already supplying the India's first large nuclear power plant under a Russian-financed 3 billion contract; and in 2014, Russia agreed to help building 10 nuclear reactors in India.

{\bf Curtailment risk:} Wind and solar curtailment may occur for several reasons including transmission congestion (or local network constraints), global oversupply and operational issues~\citep{lew2013wind}. Each type of curtailment occurs with different frequencies depending on the generation and electrical characteristics of the regional and local systems. Another example is the risk of terrorism in the congested traffic, which cannot be represented by any stochastic model.

{\bf Geopolitical implications:} Affected by the dollar, geopolitical and other factors, at the beginning of 2008 the international crude oil prices rose sharply. Another example is the Ukraine Crisis, which made Europe consider seriously adjusting its energy policy to reduce its dependence on imported energy supply.

Handling such uncertainties is a challenge.
For example, how should we modelize the risk of gas curtailment in Europe, and the evolution of oil prices ?
We discuss existing methodologies in Section \ref{decision:clas}.
Section \ref{compatools} compares them.
Section \ref{nashelec:spanas} describes our proposed approach.
In particular, Section \ref{nashelec:method} summarizes our method.
Experiments are provided in Section \ref{nashelec:real}.
Section \ref{nashelec:conc} concludes.

\section{State of the art: decision with uncertainties}\label{decision:clas}
The notations are as follows:
$K$ is the number of possible policies.
$S$ is the number of possible scenarios.
\def\K{{\cal K}}
\def\S{{\cal S}}
$R$ is the matrix of rewards and the associated reward function ($R_{k,s}=R(k,s)$), i.e. $R(k,s)$ is the reward when applying policy $k\in \K=\{1,\dots,K\}$ in case the outcome of uncertainties is $s\in\S=\{1,\dots,S\}$.
The reward function is also called a utility function or a payoff function.
A \emph{strategy} (a.k.a. policy) is a random variable $k$ with values in $\K$.
A \emph{mixed strategy} is a probability distribution of possible policies; this is the general case of a strategy.
A \emph{pure strategy} is a deterministic policy, i.e. it is a mixed strategy with probability $1$ for one element, others having probability $0$.
The \emph{exploitability} of a (deterministic or randomized) strategy $k$ is
\begin{equation}\label{eq:expl}
\left(\underset{{\small{k' stochastic}}}{\max}~\underset{s \in \{1,\dots,S\}}{\min}~\E_{k'} R(k',s)\right) - \underset{s \in \{1,\dots,S\}}{\min}~\E_kR(k,s).
\end{equation}

We refer to the choice of $s$ as Nature's choice.
This does not mean that only natural effects are involved; geopolitics and technological uncertainties are included. $k$ is chosen by us.
In fact, natural phenomena can usually be modelized with probabilities, and are included through random perturbations - they are not the point in this work - contrarily to climate change uncertainties.

\subsection{Scenario-based planning}\index{Decision under uncertainty@Scenario based planning (decision under uncertainty)}
Maybe the most usual solution consists in selecting a small set $s_1,\dots,s_M$ of possible $s$, assumed to be most realistic. Then, for each $s_j$, an optimal $k_i$ is obtained. The human then checks the matrix of the $R(k_i,s_j)$ for $i$ and $j$ in $\{1,\dots,M\}$. 
Variants of this approach are studied in scenario planning~\citep{powerplanning1,powerplanning2,artuncertainties}.
\cite{plenty2} provides examples with more than 1000 scenarios.
When optimizing the transmission network, we must take into account the future installation of power plants, for which there are many possible scenarios - in particular, the durations involved in power plant building are not necessarily larger than constants involved in big transmission lines. The scenarios involving large wind farms, or large nuclear power plants, lead to very specific constraints depending on their capacities and locations.

\subsection{Wald criterion}\label{decision:walds}\index{Decision under uncertainty@Wald (decision under uncertainty)}
The Wald criterion\citep{wald} consists in optimizing in the worst case scenario.
For a maximization problem, the \emph{Wald-value} is
\begin{equation}\label{eq:wald}
v = \underset{k\mbox{ pure strategy on }\{1,\dots,K\}}{\max}~\underset{s\in\{1,\dots,S\}}{\min}~R_{k,s},
\end{equation}
and the recommended policy is $k$ realizing the $\max$.
We choose a policy which provides the best solution (maximal reward) for the worst scenario.
Wald's maximin model provides a reward which is guaranteed in all cases. Implicitly, it assumes that Nature will make its decision in order to bother us, and, in a more subtle manner, Nature will make its decision while knowing what we are going to decide. It is hard to believe, for example, that the ultimate technological limit of photovoltaic units will be worse if we decide to do massive investments in solar power. Therefore, Wald's criterion is too conservative in many cases; hence the design of the Savage criterion.

\subsection{Savage criterion}\label{decision:savag}\index{Decision under uncertainty@Savage criterion (decision under uncertainty)}
The \emph{Savage-value}\citep{savage} is:
\begin{equation}\label{eq:savage}
v = \underset{k\mbox{ pure strategy on }\K}{\min} \underset{s \in \S}{\max}~regret(k,s), 
\end{equation}
where $regret(k,s)=\underset{k'\in \K}{\max}~(R_{k',s} - R_{k,s})$.
The Savage criterion is an application of the Wald maximin model to the regret. Contrarily to Wald's criterion, it does not focus on the worst scenario. Its interpretation is that we optimize the guaranteed loss compared to an anticipative choice (anticipative in the sense: aware of all future outcomes) of decision. On the other hand, Nature still makes its decision after us, and has access to our decision before making its decision - Nature, in this model, can still decide to reduce the technological progress of wind turbines just because we have decided to do massive investments in wind power.

\subsection{Nash equilibria}\label{decision:nash}\index{Nash equilibrium@Decision under uncertainty with Nash equilibria}

The principle of the Nash equilibrium is that contrarily to what is assumed in Wald's criterion (Eq. \ref{eq:wald}), there is no reason for Nature (the opponent) to make a decision {\em{after}} us, and to know what we have decided.  
The \emph{Nash-value} $v$ is 
\begin{equation*}\label{eq:nash}
v=\underset{k\mbox{ mixed strategy on }\K}{\max}~\underset{s \in \S}{\min}~\E_{k}R(k,s).
\end{equation*}
As a mixed strategy is used, the fact that the maximum is written before the min does not change the result~\citep{neumann1928zur}; $v$ is also equal to
$$\min_{s \mbox{ r.v. on }\S} -\max_k \E_k R(k,s).$$
where r.v. stands for ``random variable''.
The exploitability (Eq. \ref{eq:expl}) of a (possibly mixed) strategy $k$ is equivalent to
\begin{equation*}
\mbox{\emph{Nash-value}} - \underset{s \in \S}{\min}~\E_kR(k,s).
\end{equation*}
A Nash strategy is a strategy with exploitability equal to $0$.
A Nash strategy always exists; it is not necessarily unique.
A Nash equilibrium, for a finite-sum problem, is a pair of Nash strategies for us and for Nature respectively.
In the general case, a Nash strategy is not pure.
Criteria for Nash equilibria corresponds to Nature and us making decision privately, i.e. without knowing what each other will do.
In this sense, it is more intuitive than other criteria.

\subsection{Other decision tools}
Other possible tools for partially adversarial decision making are multi-objective optimization (i.e. for each $s$, there is one objective function $k\mapsto R(k,s)$) and possibilistic reasoning~\citep{possibility}. These tools rely intensively on human experts, a priori (selection of scenarios) or a posteriori (selection in the Pareto set).

\section{Comparison between various decision tools}\label{compatools}
Let us compare the various discussed policies, where $K$ is the number of possible investment policies, $S$ is the number of scenarios, $K'$ is the
number of displayed policies, $S'$ is the number of displayed policies; we provide an overview in Table \ref{decision:comp}. 
\begin{table*}[th]
\centering
\caption{\label{decision:comp}Comparison between several tools for decision under uncertainty.}
\scriptsize
\begin{tabular}{ccccc}
\multirow{3}{*}{{\bf METHOD}} &\multirow{2}{*}{{\bf EXTRACTION}} & {\bf EXTRACTION}   & \multirow{2}{*}{\bf COMPUTATIONAL} & \multirow{3}{*}{\bf INTERPRETATION} \\
       & \multirow{2}{*}{\bf OF POLICIES}   & {\bf OF CRITICAL} & \multirow{2}{*}{\bf COST} & \\
       &           & {\bf SCENARIOS}   &  &  \\
\hline\\
\multirow{2}{*}{Wald}   & \multirow{2}{*}{One}          & \multirow{2}{*}{One per policy} & \multirow{2}{*}{$K\times S$} & Nature decides later,\\
 & &  & & minimizing our reward.\\

\multirow{2}{*}{Savage} & \multirow{2}{*}{One}          & \multirow{2}{*}{One per policy} & \multirow{2}{*}{$K\times S$} & Nature decides later,\\
 & &  & & maximizing our regret.\\

\multirow{2}{*}{Nash}   & \multirow{2}{*}{Nash-optimal} & \multirow{2}{*}{Nash-optimal} & \multirow{2}{*}{$(K+S)\times\log(K+S)$} & Nature decides \\
  &  &  &  & privately, before us.\\
  
Scenarios & Handcrafted & Handcrafted & $K'\times S'$  & Human expertise  \\
\hline
\end{tabular}
\end{table*}
We see that the Nash approach (at least with the algorithms reaching the bound mentioned in the table) has a lower computational cost and some advantages in terms of modeling; Nature makes its decision privately (which means we do not know the uncertainties), but not with access to our decisions. On the other hand, its output is stochastic, which might be a drawback for users.

\section{Our proposal: NashUncertaintyDecision}\label{nashelec:spanas}
Our proposed tool is as follows:
	 (i) We use Nash equilibria, for their principled nature and (as discussed later) low computational cost in large scale settings.
	 (ii) We compute the equilibria thanks to adversarial bandit algorithms, as detailed in the next section.
	(iii) We use sparsity, for (i) improving the precision (ii) reducing the number of pure strategies in our recommendation.
The resulting algorithm has the following advantage:
\begin{itemize}
\item It is fast; this is not intuitive, but Nash equilibria, in spite of the complex theories behind this concept, can be approximated quickly, without computing the entire matrix of $R(k,s)$. A pioneering work in this direction was \cite{grigoriadis}; within logarithmic terms and dependency in the precision, the cost is roughly the square root of the size of the matrix.
\item It naturally provides a submatrix of $R(k,s)$, for the best $k$ and the most critical $s$.
\end{itemize}
We believe that such outcomes are natural tools for including in platforms for simulating large scale power systems involving huge uncertainties.

\subsection{The algorithmic technology under the hood: computing Nash equilibria with adversarial bandit algorithms}\label{nashelec:algo}
For the computational cost issue for computing Nash equilibria, there exist algorithms reaching approximate solutions much faster than the exact linear programming approach~\citep{Stengel02computeequilibria}. Some of these fast algorithms are based on the bandit formalism.
The Multi-Armed Bandit (MAB) problem~\citep{lairobbins,katehakis1987multi,auer95gambling} is a model of exploration/exploitation trade-offs, aimed at optimizing the expected payoff.
Let us define an adversarial multi-armed bandit with $K \in \N^+$ ($K>1$) arms and let $\K$ denote the set of arms.
Let $\T=\{1,\dots,T\}$ denote the set of time steps, with $T\in \N^+$ a finite time horizon.
At each time step $t\in \T$, the algorithm chooses $i_t \in \K$ and obtains a reward $R_{i_t,t}$.
The reward $R_{i_t,t}$ is a mapping $(\K, \T)\mapsto \R$.

The generic adversarial bandit is detailed in Algorithm \ref{nashelec:gsba}.
\begin{algorithm}
\caption{\label{nashelec:gsba}Generic adversarial multi-armed bandit.
The problem is described through the arm sets, the budget $T$, and most importantly the \emph{get reward} method, i.e. the mapping $(\K, \T)\mapsto \R$.}
\begin{algorithmic}[1]
\scriptsize
\Require{a time horizon (computational budget) $T \in \N^+$}
\Require{a set of arms $\K$}
\Require{a probability distribution $\pi$ on $\K$}
\For{$t\leftarrow1$ to $T$}
\State Select arm $i_t \in \K$ based upon $\pi$
\State \emph{Get reward} $R_{i_t,t}$
\State Update the probability distribution $\pi$ using $R_{i_t,t}$.
\EndFor
\end{algorithmic}
\end{algorithm}
In the case of adversarial problems, when we search for a Nash equilibrium for a reward function $(k,s)\mapsto R(k,s)$, two bandit algorithms typically play against each other. One of them is Nature, and the other plays our role. At the end, our bandit algorithm recommends a (possibly mixed) strategy over the $K$ arms. This recommended distribution is often the empirical distribution of play during the games against the Nature bandit.

Such a fast approximate solution can be provided by $Exp3$ (Exponential weights for Exploration and Exploitation)~\citep{auer2002finite} and its $Exp3.P$ variant~\citep{auer2002nonstochastic}, presented in Algorithm \ref{nashelec:exp3p}. $Exp3$ has the same efficiency as the Grigoriadis and Khachiyan method~\citep{grigoriadis} for finding approximate Nash equilibria, and can be implemented with two bandits playing one against each other, e.g. one for us and one for Nature.
$Exp3.P$ is not anytime: it requires the time horizon in order to initialize some input meta-parameters.
\begin{algorithm}[h]
\caption{\label{nashelec:exp3p}$Exp3.P$: variant of $Exp3$, proved to have a high probability bound on the weak reward.
$\eta$ and $\gamma$ are two parameters.}
\begin{algorithmic}[1]
\scriptsize
\Require{$\eta \in \R$}
\Require{$\gamma \in (0,1]$}
\Require{a time horizon (computational budget) $T \in \N^+$}
\Require{$K\in \N^+$ is the number of arms}
\State $y\leftarrow 0$
\For{$i\leftarrow 1$ to $K$}	\Comment{initialization}
	\State $\w_i\leftarrow \exp(\frac{\eta\gamma}{3}\sqrt{\frac{T}{K}})$
\EndFor
\For{$t\leftarrow 1$ to $T$}
	\For{$i\leftarrow 1$ to $K$}
		\State $p_i\leftarrow (1-\gamma)\frac{\w_i}{\sum_{j=1}^{K} \w_j} + \frac{\gamma}{K}$
	\EndFor
	\State Generate $i_t$ according to $(p_1,p_2,\dots,p_K)$
	\State \emph{Compute reward} $R_{i_t,t}$
	\For{$i\leftarrow 1$ to $K$}
		\If{$i==i_t$}
			\State $\hat R_i\leftarrow \frac{R_{i_t,t}}{p_i}$
		\Else
			\State $\hat R_i\leftarrow 0$
		\EndIf
		\State $\w_i\leftarrow \w_i \exp\left( \frac{\gamma}{3K} ( \hat R_i + \frac{\eta}{p_i\sqrt{TK}} ) \right)$
	\EndFor
\EndFor
\State\Return{probability distribution $(p_1,p_2,\dots,p_K)$}
\end{algorithmic}
\end{algorithm}
\cite{busa2010fast} optimized Adaptive Boosting (AdaBoost), a popular machine-learning meta-algorithm, by the adversarial bandit algorithm $Exp3.P$, and proposed two parametrizations of the algorithm, as detailed in Table \ref{nashelec:param}.
\cite{bubeck2012regret} proved a high probability bound on the weak reward of $Exp3.P$.
\begin{table}
\centering
\caption{\label{nashelec:param}Notations and parameters of algorithms using in the experiments.
$T$ is the horizon, i.e. simulation number.
``$+t$'' (resp. ``$+p$'') refers to the variant of $Exp3.P$ or $tExp3.P$ with theoretical (resp. practical) parametrization.
$\e$ is the precision. We use $\e=1e-6$ in our experiments.}
\scriptsize
\begin{tabular}{ccc}
\multirow{2}{*}{\bf NOTATION} & \multicolumn{2}{c}{\bf PARAMETERS OF $Exp3.P$} \\
\cline{2-3}
 & {\bf $\eta$} & {\bf $\gamma$} \\
\hline
$Exp3.P+p$ & \multirow{2}{*}{$0.3$} & \multirow{2}{*}{$0.15$} \\
$tExp3.P+p$ & & \\ 
$Exp3.P+t$ & \multirow{2}{*}{$2\sqrt{\log\frac{KT}{\e}}$} & \multirow{2}{*}{$\min(0.6,2\sqrt{\frac{3K\log(K)}{5T}})$} \\
$tExp3.P+t$ & &\\ 
\hline
\end{tabular}
\end{table}

\subsection{Another ingredient under the hood: sparsity}\label{nashelec:spars}
\cite{teytaud2011upper} proposed a truncation technique on sparse problem.
Considering the Nash equilibria for two-player finite-sum matrix games, if the Nash equilibrium of the problem is sparse, the small components of the solution can be removed and the remaining submatrix is solved exactly.
This technique can be applied to some adversarial bandit algorithm such as $Grigoriadis$' algorithm~\citep{grigoriadis}, $Exp3$~\citep{auer2002finite} or $Inf$~\citep{bubeck2009pure}. 
The properties of this sparsity technique are as follows.
	 Asymptotically in the computational budget, the convergence to the Nash equilibria is preserved~\citep{teytaud2011upper}.
	 The computation time is lower if there exists a sparse solution~\citep{teytaud2014sparse}.
	 The support of the obtained approximation has at most the same number of pure strategies and often far less~\citep{teytaud2011upper}. Essentially, we get rid of the random exploration part of the empirical distribution of play.

\subsection{Overview of our method}\label{nashelec:method}
We first give a high level view of our method, in Algorithm \ref{nashelec:SNash}. All the algorithmic challenge is hidden in the $tExp3.P$ algorithm, defined later.
\begin{algorithm}[h]
\caption{\label{nashelec:SNash}The SNash (Sparse-Nash) algorithm for solving decision under uncertainty problems.}
\begin{algorithmic}
\scriptsize
\Require{A family $\{1,\dots,K\}$ of possible decisions (investment policies).}
\Require{A family $\{1,\dots,S\}$ of scenarios.}
\Require{A mapping $(k,s)\mapsto R_{k,s}$, providing the rewards}
\State Run $tExp3.P$ on the mapping $R$, get a probability distribution on $\K$ and a probability distribution on $\S$.
\State Output $k_1,\dots,k_m$ the policies with positive probability and $s_1,\dots,s_p$ the scenarios with positive probability. Emphasize the policy with highest probability.
\State Output the matrix of $R(k_i,s_j)$ for $i\leq n$ and $j\leq p$.
\end{algorithmic}
\end{algorithm}
We now present the computation engine $tExp3.P$.
We apply the truncation technique~\citep{teytaud2011upper} to $Exp3.P$.
We present in Algorithm \ref{nashelec:texp3p} the resulting algorithm, denoted as $tExp3.P$. 
\begin{algorithm}
\caption{\label{nashelec:texp3p}$tExp3.P$, combining $Exp3.P$ and the truncation method. $\alpha$ is the truncation parameter.}
\begin{algorithmic}[1]
\scriptsize
\Require{$R_{m\times n}$, matrix defined by mapping $(i,j)\mapsto R_{i,j}$}
\Require{a time horizon (computational budget) $T \in \N^+$}
\Require{$\alpha$, truncation parameter}
\State Run $Exp3.P$ during $T$ iterations; get an approximation $(p,q)$ of the Nash equilibrium
\State $\zeta=\underset{i \in \{1,\dots,m\}}{\max} \frac{(T p_i)^{\alpha}}{T}$	\Comment{compute the threshold for $p$}
\For{$i\leftarrow 1$ to $m$}	\Comment{Truncation}
	\If{$p_i\geq \zeta$}  
		\State $p'_i=p_i$
	\Else
		\State $p'_i=0$
	\EndIf
\EndFor
\For{$i\leftarrow 1$ to $m$}
  \State $p''_i=\frac{p'_i}{\sum_{j=1}^{m} p'_j}$
\EndFor
\State $\zeta'=\underset{i \in \{1,\dots,n\}}{\max} \frac{(T q_i)^{\alpha}}{T}$	\Comment{compute the threshold for $q$}
\For{$i\leftarrow 1$ to $n$}	\Comment{Truncation}
	\If{$q_i\geq \zeta'$}  
		\State $q'_i=q_i$
	\Else
		\State $q'_i=0$
	\EndIf
\EndFor
\For{$i\leftarrow 1$ to $n$}
  \State $q''_i=\frac{q'_i}{\sum_{j=1}^{n} q'_j}$
\EndFor\\
\Return{$p''$ and $q''$ as an approximate Nash equilibrium of the problem}
\end{algorithmic}
\end{algorithm}

\section{Experiments}\label{nashelec:real}
We propose a simple model of investments in power systems.
Our model is not supposed to be realistic; it is aimed at being
easy to reproduce.

\subsection{Power investment problem}\label{nashelec:power}\index{Power investment}
We consider each investment \emph{policy}, sometimes called action or decision, a vector
$\vect k=(C,F,X,S,W,P,T,U,N,A) \in \{0,\frac12,1\}^{10}$.
A \emph{scenario} is a vector
$\vect s=(Z,WB,PB,TB,XB,UB,SB,CC,NT) \in\{0,\frac12,1\}^{9}$.
Detailed descriptions of parameters are provided in Tables \ref{nashelec:ds} and \ref{nashelec:as}.
\begin{table*}[h]
\centering
\caption{Parameters and descriptions of policy variables (vector $\vect k$) and scenario (vector $\vect s$) in power investment problem.}
\begin{subtable}[c]{0.4\textwidth}
\centering
\subcaption{\label{nashelec:ds}Parameters and descriptions of policy variables (vector $\vect k$) in power investment problem.}
\scriptsize{
\begin{tabular}{cl}
{\bf $\vect k \in\{0,\frac12,1\}$} & \multicolumn{1}{c}{\bf CORRESPONDING INVESTMENT}\\
\hline\\
C & Coal \\
F & Nuclear fission \\
X & Nuclear fusion \\
S & Supergrids \\
W & Wind power \\
P & PV units \\
T & Solar thermal \\
U & Unconventional renewable \\
N & Nanogrids \\
A & massive storage in Scandinavia \\
\end{tabular}
}
\end{subtable}
\hspace{6em}
\begin{subtable}[c]{0.4\textwidth}
\centering
\subcaption{\label{nashelec:as}Parameters and descriptions of scenario (vector $\vect s$) in power investment problem.}
\scriptsize{
\begin{tabular}{cl}
{\bf $\vect s \in\{0,\frac12,1\}$} & \multicolumn{1}{c}{\bf NATURE'S ACTION} \\
\hline\\
Z&Massive geopolitical issues \\
WB&Wind power technological breakthrough \\
PB&PV Units breakthrough \\
TB&Solar thermal breakthrough \\
XB&Fusion breakthrough \\
UB&Unconventional renewable breakthrough \\
SB&Local storage breakthrough \\
CC&Climate change disaster \\
NT&Nuclear terrorism \\
\end{tabular}
}
\end{subtable}
\end{table*}

Let $\S$ be the set of possible scenarios and $\K$ be the set of possible policies.
The utility function $R$ is a mapping $(\K,\S)\mapsto \R$.
Given decision $\vect k\in\K$ and scenario $\vect s\in\S$, a reward can be computed by 
\begin{align*}
&R(\vect k,\vect s)=\frac23(1+rand)\cdot(N(1-Z)/5\\
&-cost\cdot (N+U+T+P+W+S+X+F+C)\\
&+7XB\cdot X +W(1+WB)(SB+\sqrt{S})/2\\
&+3P(PB+SB) -4C\cdot CC -F\cdot NT\\
&+S(1-Z) +P\cdot Z +U\cdot UB\\
&+T\cdot S\cdot(1+TB-SB/2)\\
&-F\cdot NT+A\cdot(1+W+P-2SB)).
\end{align*}
where $cost$ is a meta-parameter.
This provides a reward function $R(\vect k,\vect s)$, with which we can build a matrix $R$ of rewards. However, with a ternary discretization for each variable we get a huge matrix, that we will not construct explicitly - more precisely, it would be impossible to construct it explicitly with a real problem involving hours of computation for each $R(\vect k,\vect s)$. Fortunately, approximate algorithms can solve Nash equilibria with precision $\e$ with $O(K\log(K)/\e^2)$ requests to the reward function, i.e. far less than the quadratic computation time $K^2$ needed for reading all entries in the matrix.
We do experiments on this investment problem and apply the algorithms described in Table \ref{nashelec:param}.
We consider policies and scenarios in discrete domains:
$\K = \{0,\frac12,1\}^{10}$,~ $\S= \{0,\frac12,1\}^{9}$.
The reward matrix $R_{3^{10} \times 3^9}$ can be defined by
$\forall i \in \{1,\dots,3^{10}\},\ \forall j \in \{1,\dots,3^{9}\},\ R_{i,j}= R(\vect k_i,\vect s_j)$, but the reward is noisy as previously mentioned,
where $\vect k_i$ denotes the $i^{th}$ policy in $\K$ and $\vect s_i$ denotes the $j^{th}$ scenario in $\S$.
Thus, each line of the matrix is a possible policy and each column is a scenario, $R_{i,j}$ is the reward obtained by apply the policy $\vect k_i$ to the scenario $\vect s_j$.
Experiments are performed for different numbers of time steps in the bandit algorithms, i.e. we consider $T$ simulations for each $T \in \{1,2,8,10,32,128,512,2048\} \times \lceil{3^{10}/10}\rceil$.
Thus when playing with the ``theoretical'' parametrization, for each $T$, the input meta-parameters $\eta$ and $\gamma$ are different, as they depend on the budget $T$.
In the entire paper, when we show an expected reward $R(\vect k,\vect s)$ for some $\vect s$ and for $\vect k$ learned by one of our methods, we refer to $10000$ trials; $R(\vect k,\vect s)$ are played for $10000$ randomly drawn pairs $(\vect k_{i_n},\vect s_{j_n})$ i.i.d. according to the random variables $i_n$ and $j_n$ proposed by the considered policies.
The performance is the average reward of these $10000$ trials $R(\vect k_{i_1},\vect s_{j_1}),\dots,R(\vect k_{i_{10000}},\vect s_{j_{10000}})$. There is an additional averaging, over learning. Namely, each learning (i.e. the sequence of $Exp3$ iteration for approximating a Nash equilibrium) is repeated $100$ times.
The meta-parameter $cost$ is set to $1$ in our experiments.

We use the parametrizations of variants of $Exp3.P$ presented in Table \ref{nashelec:param}.
\cite{teytaud2011upper} proposed $\alpha=0.7$ as truncation parameter in truncated $Exp3.P$ and \cite{teytaud2014sparse} used the same value.
The sparsity level, as well as the performance, are given in Table \ref{nashelec:sparse}.
We validate the good performance of $\alpha=0.7$. However, the sparsity is better with higher values - but these higher values do not always provide better results than the original non-sparse bandit.

We observe that when the number of simulations is bigger than the cardinality of the search domain, i.e. the number of possible pure policies, then $\alpha\simeq 0.9$ leads to better empirical mean reward against the uniform policy. Values between 0.5 and 1 are the best ones.
 When learning with few simulations ($5905=\lceil{K/10}\rceil$), the non-truncated solutions and non-sparse solutions are as weak as a random strategy.
 Along with the increment of simulation times, the non-truncated solutions and non-sparse solutions become stronger, but still weaker than the truncated solutions.
When we use the truncation, we get significant mean reward even with a small horizon, i.e. the $tExp3.P+t$ succeeds in finding better and ``purer' policies than $Exp.3$.

\subsubsection{The parameters of $Exp3.P+t$}\label{nashelec:eta}

\begin{table*}[htpb]
\centering
\scriptsize
\caption{\label{nashelec:sparse}In these tables, the result is averaged over $100$ independent learnings. The standard deviation is shown after $\pm$. ``simul.'' refers to ``simulation number'', i.e. horizon.
{\bf{Top: }}Average sparsity level (over $3^{10}=59049$ arms), i.e. number of pure policies in the support of the obtained approximation, of solutions provided by $Exp3.P+t$ in power investment problem. ``non-truncated'' means that all elements of the solution provided are below the threshold $\zeta$ (line 2 in Algorithm \ref{nashelec:texp3p}); ``non-sparse'' means that all elements of the solution provided are above the threshold $\zeta$. In both cases, we play with the original solution provided by $Exp3.P+t$.
{\bf{Middle: }}Empirical mean reward obtained using different truncation parameter $\alpha$.
{\bf{Bottom: }}Exploitability using different truncation parameter $\alpha$ for solutions provided by $Exp3.P+t$ in power investment problem.  In this table, we exclude the solutions which can not be truncated, i.e. all elements are below or all elements are above the truncation threshold $\zeta$.
If at least one solution is excluded, we show between parenthesis the number of solutions excluded from the $100$ learning.  When $\alpha=1.0$, $\zeta={\max}_{i\in \{1,\dots,K\}} p_i$. Thus, the remaining policy (policies), after truncation, is the element with the highest frequency (i.e., except in the rare case of a tie, only one arm).}
\tiny
\resizebox{\textwidth}{!}{%
\begin{tabular}{ccccccc}
\multirow{2}{*}{$\alpha$} & \multicolumn{6}{c}{\bf{AVERAGE SPARSITY LEVEL OVER $3^{10}=59049$ ARMS}}\\
\cline{2-7}
& $8\lceil K/10 \rceil$ simul.& $K$ simul.& $32\lceil K/10 \rceil$ simul.& $128\lceil K/10 \rceil$ simul.& $512\lceil K/10 \rceil$ simul.& $2048\lceil K/10 \rceil$ simul.\\ 
\hline\\
0.1 & 1873.70 $\pm$ 10.87& 4621.87 $\pm$ 28.34& non-truncated& non-truncated& non-truncated& non-truncated\\
0.3 & 491.53 $\pm$ 7.74 & 955.32 $\pm$ 13.78 & 12140.13 $\pm$ 234.46 & 7577.95 $\pm$ 154.37 & 710.45 $\pm$ 11.28 & 320.43 $\pm$ 2.91\\
\bf{0.5} & 126.18 $\pm$ 3.71& 216.63 $\pm$ 5.53 & 1502.24 $\pm$ 33.85 & 687.42 $\pm$ 19.37 & 33.01 $\pm$ 1.14 & 10.16 $\pm$ 0.27\\
\bf{0.7} & 24.80 $\pm$ 1.23 & 36.69 $\pm$ 1.69 & 168.02 $\pm$ 6.94 & 63.04 $\pm$ 2.49 & 6.57 $\pm$ 0.27 & 2.59 $\pm$ 0.11\\
\bf{0.9} & 3.54 $\pm$ 0.23 & 3.73 $\pm$ 0.26 & 7.35 $\pm$ 0.49 & 5.12 $\pm$ 0.29 & 1.93 $\pm$ 0.09 & 1.17 $\pm$ 0.04\\
\hline\\
\multirow{2}{*}{$\alpha$} & \multicolumn{6}{c}{\bf{EMPIRICAL MEAN REWARD AGAINST PURE STRATEGIES}}\\
\cline{2-7}
& $8\lceil K/10 \rceil$ simul.& $K$ simul.& $32\lceil K/10 \rceil$ simul.& $128\lceil K/10 \rceil$ simul.& $512\lceil K/10 \rceil$ simul.& $2048\lceil K/10 \rceil$ simul.\\ 
\hline\\
0.1 & 2.595 $\pm$ .006 & 2.174 $\pm$ .006 & -.029 $\pm$ .004 & 1.050 $\pm$ .004 & 2.184 $\pm$ .005 & 4.105 $\pm$ .006 \\
0.3 & 3.299 $\pm$ .010 & 3.090 $\pm$ .009 & 2.195 $\pm$ .017 & 3.892 $\pm$ .018 & 6.555 $\pm$ .008 & 6.822 $\pm$ .004 \\
\bf{0.5} & 3.896 $\pm$ .016 & 3.779 $\pm$ .015 & 3.592 $\pm$ .016 & 5.275 $\pm$ .020 & 6.741 $\pm$ .007 & 6.853 $\pm$ .004 \\
\bf{0.7} & 4.501 $\pm$ .030 & 4.454 $\pm$ .027 & 4.674 $\pm$ .022 & 6.101 $\pm$ .016 & 6.777 $\pm$ .007 & 6.858 $\pm$ .005 \\
\bf{0.9} & {\bf 5.021 $\pm$ .058} & {\bf 5.149 $\pm$ .062} & 5.703 $\pm$ .040 & {\bf 6.536 $\pm$ .017} & {\bf 6.813 $\pm$ .007} & {\bf 6.873 $\pm$ .004} \\
\bf{Pure} & 4.853 $\pm$ .158 & 5.027 $\pm$ .143 & {\bf 5.709 $\pm$ .101} & 6.137 $\pm$ .163 & 6.413 $\pm$ .136 & 6.844 $\pm$ .028 \\
\hline\\
\multirow{2}{*}{$\alpha$} & \multicolumn{6}{c}{\bf{EXPLOITABILITY INDICATOR: WORST SCORE AGAINST PURE STRATEGIES}}\\
\cline{2-7}
& $8\lceil K/10 \rceil$ simul.& $K$ simul.& $32\lceil K/10 \rceil$ simul.& $128\lceil K/10 \rceil$ simul.& $512\lceil K/10 \rceil$ simul.& $2048\lceil K/10 \rceil$ simul.\\ 
\hline\\
\bf{0.5} & $-5.560 \pm 0.070$ & $-5.693 \pm 0.058$ & $-5.725 \pm 0.060$ & $-3.479 \pm 0.061$ & $-0.576 \pm 0.041$ & $0.056 \pm 0.024$\\
\bf{0.7} & $-4.028 \pm 0.094$ & $-4.132 \pm 0.094$ & $-4.038 \pm 0.074$ & $-1.243 \pm 0.032$ & $0.010 \pm 0.018$ & $0.268 \pm 0.011$\\
\bf{0.9} & $-2.012 \pm 0.107$ & $-1.859 \pm 0.115$ & $-1.369 \pm 0.081$ & $-0.195 \pm 0.028$ & $0.272 \pm 0.011$ & $0.330 \pm 0.003$\\
\bf{Pure} & ${\bf -0.938 \pm 0.078}$ & ${\bf -0.971 \pm 0.092}$ & ${\bf -0.455 \pm 0.060}$ & ${\bf 0.182 \pm 0.021}$ & ${\bf 0.323 \pm 0.005}$ & \bf{0.333 $\pm$ 0.000}\\
\hline
\end{tabular}
}
\end{table*}

When learning with few simulations ($5905=\lceil{K/10}\rceil$), the non-truncated solutions and non-sparse solutions are as weak as a random strategy.
Along with the increment of simulation times, the non-truncated solutions and non-sparse solutions become stronger, but still weaker than the truncated solutions.
Sparsity level ``0.01'' means that one and only one solution of the $100$ learnings has one element above the threshold $\zeta$, the other $99$ solutions of the $99$ learnings have no element above the threshold $\zeta$.
This situation is not far from the non-truncated or non-sparse case.
If the solution is sparse, we get a better empirical mean reward even with a small horizon, i.e. the $tExp3.P+t$ succeeds in finding better pure policies.

We see that truncated algorithms outperform their non-truncated counterparts, in particular, truncation clearly shows its strength when the number of simulations is small in front of the size of search domain.

\subsection{A modified power investment problem}\label{nashelec:newr}\index{Power investment}
Now we modify the reward function as follows:
\begin{align*}
R'(\vect k,\vect s)=&R(\vect k,\vect s)+\mathbf{c\cdot((X==XB)}\\
&\mathbf{+(C~=CC)+(NT~=F)+(P==PB))}.
\end{align*}
where $cost$ and $c$ are meta-parameters.

As presented in the previous section, we can build a matrix $R'$ with the reward function $R'(\vect k,\vect s)$.
We do experiments on this modified investment problem and apply the algorithms described in Table \ref{nashelec:param}.
We consider policies and scenarios in discrete domains as used in the previous section.
The meta-parameters $cost$ is set to $1$ and $c$ is set to $\{1,2,\dots,10\}$ in our experiments.
The reward matrix is normalized in the experiments.
\begin{table*}[htpb]
\centering
\scriptsize
\caption{\label{nashelec:c1}Results for reward matrix $R'$ computed with $c=1$. 
In these tables, the result is the average value of $100$ learnings.
The standard deviation is shown after $\pm$.
``NT'' means that the truncation technique is not applied; ``non-sparse'' means that all elements of the solution provided are above the threshold $\zeta$. 
{\bf{Top: }}Average sparsity level (over $3^{10}=59049$ arms), i.e. number of pure policies in the support of the obtained approximation, of solutions provided by $Exp3.P+t$ in power investment problem.
{\bf{Middle: }}Proxy exploitability (to be maximized) using different truncation parameter $\alpha$ for solutions provided by $Exp3.P+t$ in power investment problem.
The proxy exploitability is the difference between the best robust score in the table, minus the robust score.
{\bf{Bottom: }}Robust score (to be minimized) using different truncation parameter $\alpha$ for solutions provided by $Exp3.P+t$ in power investment problem.
The robust score is the worst of the scores against pure policies.
}
\begin{tabular}{ccccccc}
\multirow{2}{*}{$\alpha$} & \multicolumn{6}{c}{\bf{AVERAGE SPARSITY LEVEL OVER $3^{10} = 59049$ ARMS}}\\
\cline{2-7}
& $T=K$ & $T=10K$ & $T=50K$ & $T=100K$ & $T=500K$ & $T=1000K$ \\ 
\hline
0.1 & 13804.380 $\pm$ 52.015 & non-sparse & non-sparse & non-sparse & non-sparse & non-sparse\\
0.3 & 2810.120 $\pm$ 59.083 & non-sparse & non-sparse & non-sparse & non-sparse & non-sparse\\
0.5 & 395.920 $\pm$ 15.835 & non-sparse & non-sparse & 59048.960 $\pm$ 196.946 & 49819.430 $\pm$ 195.016 & non-sparse\\
0.7 & 43.230 $\pm$ 2.624 & 58925.340 $\pm$ 26.821 & 55383.140 $\pm$ 150.057 & 46000.020 $\pm$ 277.653 & 9065.180 $\pm$ 159.610 & non-sparse\\
0.9 & 3.600 $\pm$ 0.260 & 992.940 $\pm$ 64.474 & 796.500 $\pm$ 41.724 & 503.600 $\pm$ 24.927 & 97.670 $\pm$ 5.445 & 52632.820 $\pm$ 522.505\\
0.99 & 1.110 $\pm$ 0.031 & 2.250 $\pm$ 0.171 & 2.500  $\pm$ 0.180 & 2.310 $\pm$ 0.156 & 1.790 $\pm$ 0.121 & 6.700 $\pm$ 0.612\\
\hline\\
\multirow{2}{*}{$\alpha$} & \multicolumn{6}{c}{\bf{PROXY EXPLOITABILITY}}\\
\cline{2-7}
& $T=K$ & $T=10K$ & $T=50K$ & $T=100K$ & $T=500K$ & $T=1000K$\\
\hline
NT & 4.922e-01 $\pm$ 5.649e-07 & 4.928e-01 $\pm$ 1.787e-06 & 4.956e-01 $\pm$ 4.016e-06 & 4.991e-01 $\pm$ 5.892e-06 & 5.221e-01 $\pm$ 1.404e-05 & 4.938e-01 $\pm$ 1.687e-06\\
0.1 & 4.948e-01 $\pm$ 5.739e-05 & 4.928e-01 $\pm$ 1.787e-06 & 4.956e-01 $\pm$ 4.016e-06 & 4.991e-01 $\pm$ 5.892e-06 & 5.221e-01 $\pm$ 1.404e-05 & 4.938e-01 $\pm$ 1.687e-06\\
0.3 & 5.004e-01 $\pm$ 1.397e-04 & 4.928e-01 $\pm$ 1.787e-06 & 4.956e-01 $\pm$ 4.016e-06 & 4.991e-01 $\pm$ 5.892e-06 & 5.221e-01 $\pm$ 1.404e-05 & 4.938e-01 $\pm$ 1.687e-06\\
0.5 & {\bf{5.059e-01 $\pm$ 2.272e-04}} & 4.928e-01 $\pm$ 1.787e-06 & 4.956e-01 $\pm$ 4.016e-06 & 4.991e-01 $\pm$ 5.891e-06 & 5.242e-01 $\pm$ 5.491e-05 & 4.938e-01 $\pm$ 1.687e-06\\
0.7 & 5.054e-01 $\pm$ 1.327e-03 & 4.928e-01 $\pm$ 3.835e-06 & 4.965e-01 $\pm$ 3.896e-05 & 5.031e-01 $\pm$ 1.016e-04 & 5.317e-01 $\pm$ 9.573e-05 & 4.938e-01 $\pm$ 1.687e-06\\
0.9 & 4.281e-01 $\pm$ 6.926e-03 & {\bf{5.137e-01 $\pm$ 4.199e-04}} & {\bf{5.151e-01 $\pm$ 5.007e-04}} & {\bf{5.140e-01 $\pm$ 4.965e-04}} & {\bf{5.487e-01 $\pm$ 9.413e-04}} & 4.960e-01 $\pm$ 1.828e-04\\
0.99& 3.634e-01 $\pm$ 8.191e-03 & 4.357e-01 $\pm$ 6.873e-03 & 4.612e-01 $\pm$ 5.380e-03 & 4.683e-01 $\pm$ 4.834e-03 & 5.242e-01 $\pm$ 3.302e-03 & {\bf{5.390e-01 $\pm$ 3.167e-03}}\\
Pure& 3.505e-01 $\pm$ 7.842e-03 & 3.946e-01 $\pm$ 7.181e-03 & 4.287e-01 $\pm$ 6.203e-03 & 4.489e-01 $\pm$ 5.410e-03 & 5.143e-01 $\pm$ 3.597e-03 & 4.837e-01 $\pm$ 5.558e-03\\  
\hline\\
\multirow{2}{*}{$\alpha$} & \multicolumn{6}{c}{\bf{ROBUST SCORE}}\\
\cline{2-7}
& $T=K$ & $T=10K$ & $T=50K$ & $T=100K$ & $T=500K$ & $T=1000K$ \\
\hline
NT  & 1.369e-02  & 2.092e-02  & 1.946e-02  & 1.492e-02  & 2.669e-02  & 4.525e-02\\                                         
0.1  & 1.109e-02  & 2.092e-02  & 1.946e-02  & 1.492e-02  & 2.669e-02  & 4.525e-02\\
0.3  & 5.485e-03  & 2.092e-02  & 1.946e-02  & 1.492e-02  & 2.669e-02  & 4.525e-02\\
0.5  & {\bf{0.000e+00}}  & 2.092e-02  & 1.946e-02  & 1.492e-02  & 2.454e-02  & 4.525e-02\\
0.7  & 4.328e-04  & 2.091e-02  & 1.854e-02  & 1.083e-02  & 1.705e-02  & 4.525e-02\\
0.9  & 7.778e-02  & {\bf{0.000e+00}}  & {\bf{0.000e+00}}  & {\bf{0.000e+00}}  & {\bf{0.000e+00}}  & 4.304e-02\\
0.99  & 1.425e-01  & 7.806e-02  & 5.385e-02  & 4.564e-02  & 2.456e-02  & {\bf{0.000e+00}}\\
Pure  & 1.554e-01  & 1.191e-01  & 8.638e-02  & 6.503e-02  & 3.443e-02  & 5.537e-02\\
\hline
\end{tabular}
\end{table*}

\begin{table*}[b]
\centering
\scriptsize
\caption{\label{nashelec:c10}Results for reward matrix $R'$ computed with $c=10$. 
In these tables, the result is the average value of $100$ learnings.
The standard deviation is shown after $\pm$.
``NT'' means that the truncation technique is not applied; ``non-sparse'' means that all elements of the solution provided are above the threshold       $\zeta$. 
{\bf{Top: }}Average sparsity level (over $3^{10}=59049$ arms), i.e. number of pure policies in the support of the obtained approximation, of solutions   provided by $Exp3.P+t$ in power investment problem.
{\bf{Middle: }}Proxy exploitability (to be maximized) using different truncation parameter $\alpha$ for solutions provided by $Exp3.P+t$ in power        investment problem.
The proxy exploitability is the difference between the best robust score in the table, minus the robust score.
{\bf{Bottom: }}Robust score (to be minimized) using different truncation parameter $\alpha$ for solutions provided by $Exp3.P+t$ in power investment     problem.
The robust score is the worst of the scores against pure policies.
}
\begin{tabular}{c|cccccc}
\multirow{2}{*}{$\alpha$} & \multicolumn{6}{c}{\bf{AVERAGE SPARSITY LEVEL OVER $3^{10} = 59049$ ARMS}}\\
\cline{2-7}
& $T=K$ & $T=10K$ & $T=50K$ & $T=100K$ & $T=500K$ & $T=1000K$ \\ 
\hline
0.1 & 6394.625 $\pm$ 84.308 & non-sparse & non-sparse & non-sparse & non-sparse & non-sparse\\
0.3 & 1337.896 $\pm$ 40.491 & non-sparse & non-sparse & non-sparse & non-sparse & non-sparse\\
0.5 & 206.146 $\pm$ 12.647 & non-sparse & non-sparse & non-sparse & non-sparse & non-sparse\\
0.7 & 25.563 $\pm$ 2.045 & non-sparse & non-sparse & non-sparse & 59048.750 $\pm$ 0.250 & non-sparse\\
0.9 & 3.729 $\pm$ 0.353 & 42616.313 $\pm$ 1476.644 & 47581 $\pm$ 1015.506 & 38361.182 $\pm$ 1091.373 & 4510.125 $\pm$ 726.595 & 58323.125 $\pm$ 157.971\\
0.99 & 1.208 $\pm$ 0.072 & 4.479 $\pm$ 0.575 & 5.333 $\pm$ 0.565 & 6.000 $\pm$ 0.969 & 2.875 $\pm$ 1.076 & 8.500 $\pm$ 2.204\\
\hline\\
\multirow{2}{*}{$\alpha$} & \multicolumn{6}{c}{\bf{PROXY EXPLOITABILITY}}\\
\cline{2-7}
& $T=K$ & $T=10K$ & $T=50K$ & $T=100K$ & $T=500K$ & $T=1000K$ \\
\hline
NT & 1.151e-01 $\pm$ 6.772e-08 & 1.151e-01 $\pm$ 2.175e-07 & 1.153e-01 $\pm$ 3.707e-07 & 1.154e-01 $\pm$ 5.797e-07 & 1.167e-01 $\pm$ 2.046e-06 & 1.152e-01 $\pm$ 1.297e-06\\                                                                                     
0.1 & 1.158e-01 $\pm$ 1.843e-05 & 1.151e-01 $\pm$ 2.175e-07 & 1.153e-01 $\pm$ 3.707e-07 & 1.154e-01 $\pm$ 6.019e-07 & 1.167e-01 $\pm$ 2.046e-06 & 1.152e-01 $\pm$ 1.297e-06\\
0.3 & 1.160e-01 $\pm$ 3.441e-05 & 1.151e-01 $\pm$ 2.175e-07 & 1.153e-01 $\pm$ 3.707e-07 & 1.154e-01 $\pm$ 6.019e-07 & 1.167e-01 $\pm$ 2.046e-06 & 1.152e-01 $\pm$ 1.297e-06\\
0.5 & {\bf{1.166e-01 $\pm$ 9.751e-05}} & 1.151e-01 $\pm$ 2.175e-07 & 1.153e-01 $\pm$ 3.707e-07 & 1.154e-01 $\pm$ 5.797e-07 & 1.167e-01 $\pm$ 2.046e-06 & 1.152e-01 $\pm$ 1.297e-06\\
0.7 & 1.165e-01 $\pm$ 6.176e-04 & 1.151e-01 $\pm$ 2.175e-07 & 1.153e-01 $\pm$ 3.707e-07 & 1.154e-01 $\pm$ 5.797e-07 & 1.167e-01 $\pm$ 2.051e-06 & 1.152e-01 $\pm$ 1.297e-06\\
0.9 & 1.068e-01 $\pm$ 3.176e-03 & {\bf{1.156e-01 $\pm$ 4.586e-05}} & 1.160e-01 $\pm$ 6.348e-05 & 1.172e-01 $\pm$ 9.722e-05 & {\bf{1.266e-01 $\pm$ 3.829e-04}} & 1.152e-01 $\pm$ 4.288e-06\\
0.99& 8.423e-02 $\pm$ 3.118e-03 & 1.119e-01 $\pm$ 2.316e-03 & {\bf{1.189e-01 $\pm$ 1.888e-03}} & {\bf{1.202e-01 $\pm$ 2.101e-03}} & 1.145e-01 $\pm$ 4.519e-03 & {\bf{1.186e-01 $\pm$ 7.684e-04}}\\
Pure & 7.810e-02 $\pm$ 2.570e-03 & 8.354e-02 $\pm$ 2.710e-03 & 9.327e-02 $\pm$ 2.202e-03 & 9.658e-02 $\pm$ 2.097e-03 & 1.120e-01 $\pm$ 3.625e-03 & 8.755e-02 $\pm$ 6.497e-03\\
\hline\\
\multirow{2}{*}{$\alpha$} & \multicolumn{6}{c}{\bf{ROBUST SCORE}}\\
\cline{2-7}
& $T=K$ & $T=10K$ & $T=50K$ & $T=100K$ & $T=500K$ & $T=1000K$\\
\hline
NT &  1.494e-03  & 4.594e-04  & 3.592e-03  & 4.772e-03  & 9.903e-03  & 3.388e-03\\                                       
0.1 &  7.727e-04  & 4.594e-04  & 3.592e-03  & 4.772e-03  & 9.903e-03  & 3.388e-03\\
0.3 &  5.838e-04  & 4.594e-04  & 3.592e-03  & 4.772e-03  & 9.903e-03  & 3.388e-03\\
0.5 &  {\bf{0.000e+00}}  & 4.594e-04  & 3.592e-03  & 4.772e-03  & 9.903e-03  & 3.388e-03\\
0.7 &  9.391e-05  & 4.594e-04  & 3.592e-03  & 4.772e-03  & 9.903e-03  & 3.388e-03\\
0.9 &  9.758e-03  & {\bf{0.000e+00}}  & 2.860e-03  & 2.992e-03  & {\bf{0.000e+00}}  & 3.371e-03\\
0.99 &  3.236e-02  & 3.647e-03  & {\bf{0.000e+00}}  & {\bf{0.000e+00}}  & 1.211e-02  & {\bf{0.000e+00}}\\
Pure &  3.848e-02  & 3.204e-02  & 2.559e-02  & 2.362e-02  & 1.463e-02  & 3.103e-02\\
\hline
\end{tabular}
\end{table*}

We present the results with $c=1$ and $c=10$ in Tables \ref{nashelec:c1} and \ref{nashelec:c10}:
in both testcases, $\alpha=0.9$ does not provide good results when $T=K$, however $\alpha=0.7$ (recommended by previous works) is always better than the baseline, to which the truncation technique is not applied;
for the testcase with $c=1$, $\alpha=0.9$ outperforms the other values of $\alpha$ at most of time;
when the budget is big, $\alpha=0.99$ provides better results.

\section{Conclusion: Nash-methods have computational and modeling advantages for decision making under uncertainties}\label{nashelec:conc}
We propose a new criterion (based on Nash equilibria) and a new methodology (based on adversarial bandits + sparsity) for decision making with uncertainty.
Technically speaking, we tuned a parameter-free adversarial bandit algorithm $tExp3.P+t$, for large scale problems, efficient in terms of performance itself, and also in terms of sparsity. $tExp3.P+t$ performed better than $tExp3.P$ without truncation.
Moreover, $tExp3.P+t$ with truncation parameter $\alpha=0.7$, which is theoretically guaranteed ~\citep{teytaud2011upper}, got stable performance in the experiments. 

From a user point of view, we propose a tool with the following advantages:
	(i) Natural extraction of interesting policies and critical scenarios. However, we point out that $\alpha=.7$ provides stable (and proved) results, but the extracted submatrix becomes easily readable (small enough) with larger values of $\alpha$.
	(ii) Faster computational cost than the Wald or Savage classical methodologies.
Our methodology only requires a mapping $R:(k,s)\mapsto R(k,s)$, which computes the outcome if we use the policy $k$ and the outcome is the scenario $s$. Multiple objective functions can be handled: if we have two objectives (e.g. economy and greenhouse gas pollution), we can just duplicate the scenarios, one for which the criterion is economy, and one for which the criterion is greenhouse gas.
Given a problem, the algorithm will display a matrix of rewards for different policies and for several scenarios (including, by the trick above, several criteria such as particular matter, greenhouse, and cost).

As a summary, we get a fast criterion, faster than Wald's or Savage's criteria, with a natural interpretation. The algorithm naturally provides a matrix of results, namely the matrix of outcomes in the most interesting decisions and for the most critical scenarios. These decisions and scenarios are also equipped with a ranking.
\balance
\bibliographystyle{plainnat} 

\end{document}